\documentclass[aps,prd,preprint,nofootinbib,superscriptaddress]{revtex4-2}
\usepackage[a4paper,top=2.5cm,bottom=2.5cm,left=3.0cm,right=2.5cm]{geometry}
\usepackage{amsmath,amssymb,amsfonts}
\usepackage{bm}
\usepackage{graphicx}
\usepackage{booktabs}
\usepackage{hyperref}
\usepackage{xcolor}
\usepackage{multirow}
\begin{document}

\title{Resonance--continuum interference with a possible tensor $X(6900)$ in radiative double-charmonium production at $\sqrt{s}=10.58~{\rm GeV}$}

	\author{Meng-Kun Jia $^{(a)}$}
	\author{Yi-Jie Li $^{(a)}$}
	\author{Guang-Zhi Xu $^{(a)}$}
	\email{xuguangzhi@lnu.edu.cn}
	\author{Kui-Yong Liu $^{(b,a)}$}
	\email{liukuiyong@lnu.edu.cn}
	\affiliation{ {\footnotesize (a)~School of Physics, Liaoning University, Shenyang 110036, China}\\
		{\footnotesize (b)~School of Physics and Electronic Technology, Liaoning Normal University, Dalian 116029, China}}

\begin{abstract}

We study radiative double-charmonium production, $e^+e^-\to HH\gamma$ with $H=J/\psi$ and $\eta_c$, at $\sqrt{s}=10.58~{\rm GeV}$, including a possible tensor contribution from the fully charmed state $X(6900)$. The nonresonant amplitudes are calculated within the color-singlet non-relativistic QCD framework, while the $J^{PC}=2^{++}$ resonance is combined coherently with the continuum. We find that continuum--resonance interference significantly modifies the invariant-mass distributions near $M_{HH}\simeq M_X$. The $J/\psi J/\psi\gamma$ channel exhibits a pronounced phase-dependent resonant structure, with a near-resonance cross section of $0.038$--$0.111~{\rm fb}$. In contrast, the $\eta_c\eta_c\gamma$ channel is continuum dominated and shows only a localized interference effect with an extremely suppressed production rate, rendering its experimental observation extremely challenging. These channel-dependent features arise from the different effective decay structures and demonstrate the importance of an amplitude-level treatment of the $X(6900)$ contribution.

\end{abstract}

\maketitle

\section{Introduction}

The spectroscopy of fully heavy exotic hadrons provides a useful setting for investigating multiquark dynamics in QCD~\cite{Esposito:2016noz,Chen:2016qju,Lebed:2016hpi,Guo:2017jvc}. Unlike multiquark hidden-charm systems containing light quarks, fully charmed states are not directly governed by long-range light-meson exchange and are therefore expected to be particularly sensitive to short-distance heavy-quark interactions, nontrivial color configurations, and coupled-channel dynamics. The observation of several structures in the di-$J/\psi$ invariant-mass spectrum has consequently established the $cc\bar c\bar c$ sector as an important testing ground for fully heavy multiquark dynamics~\cite{LHCb:2020bwg,ATLAS:2023bft,CMS:2023owd}.

A prominent structure in this spectrum is the enhancement near $6.9~\mathrm{GeV}$, commonly referred to as $X(6900)$. The LHCb Collaboration first reported a narrow enhancement in this mass region, together with a broad structure above the di-$J/\psi$ threshold~\cite{LHCb:2020bwg}. Subsequent measurements by the ATLAS and CMS collaborations provided further evidence for nontrivial structures in di-charmonium final states~\cite{ATLAS:2023bft,CMS:2023owd}. In particular, the CMS analysis indicated that interference among different amplitudes can play an important role in shaping the measured $J/\psi J/\psi$ spectrum.

Various interpretations have been proposed, including compact fully charmed tetraquarks, QCD sum-rule states, coupled-channel effects, threshold rescattering, and more general line-shape mechanisms~\cite{Wang:2020tpt,Dong:2020nwy,Guo:2020pvt,Liang:2021fzr,Zhuang:2021pci,Wang:2022xja}. The present work does not attempt to distinguish among these microscopic pictures. Instead, we focus on a more general issue suggested by the measured spectra: a resonance contribution need not appear as an isolated Breit--Wigner peak, but may be substantially modified by its interference with the nonresonant continuum. A consistent description of the line shape should therefore combine the resonant and nonresonant contributions coherently at the amplitude level.

Most studies of $X(6900)$ have so far concentrated on hadron-collider production, where the measured di-$J/\psi$ spectrum may receive contributions from single-parton scattering, double-parton scattering, and nonresonant continuum production~\cite{Kom:2011bd,Borschensky:2016nkv,Maciula:2020wri}, together with final-state interactions and coupled-channel effects~\cite{Dong:2020nwy,Guo:2020pvt,Liang:2021fzr}. It is therefore worthwhile to examine the possible impact of $X(6900)$ in a complementary production environment in which the relevant exclusive final state and its continuum contribution can be treated explicitly.

In this work, we study exclusive radiative double-charmonium production in $e^+e^-$ annihilation,
\begin{equation}
	e^-(l_1)+e^+(l_2)
	\to \gamma^\ast(k)
	\to H(p_3)+H(p_4)+\gamma(p_5),
	\qquad
	H=J/\psi,\eta_c ,
	\label{eq:all}
\end{equation}
The emission of the final-state photon allows the double-charmonium subsystem to span a continuous range of invariant masses even at a fixed $e^+e^-$ center-of-mass energy. Its invariant mass $M_{HH}^2=(p_3+p_4)^2$,can therefore be reconstructed directly and provides the natural variable for studying a possible intermediate resonance. The radiative topology also permits the production of a positive-$C$ double-charmonium subsystem through an intermediate state such as a tensor resonance with $J^{PC}=2^{++}$.

The resonance mechanism considered here is
\begin{equation}
	\gamma^\ast(k)\to \gamma(p_5)X(6900),
	\qquad
	X(6900)\to H(p_3)H(p_4),
\end{equation}
with $H=J/\psi$ or $\eta_c$. The resonance thus appears as a structure in the $M_{HH}$ distribution rather than in the fixed $e^+e^-$ center-of-mass energy spectrum. Radiative production consequently makes it possible to examine the double-charmonium line shape without scanning the collider energy across the resonance region.

Previous studies have investigated the exclusive radiative production of fully charmed tetraquarks at $B$ factories~\cite{Feng:2020qee}. The present analysis differs in that the nonresonant $HH\gamma$ continuum is included explicitly and combined coherently with the resonance amplitude. We take the $J^{PC}=2^{++}$ assignment as a representative benchmark, without specifying the microscopic structure of the state. The full contribution is described at the amplitude level, allowing for an unknown relative phase between the continuum and the $X(6900)$ resonance. The continuum contribution is calculated within the color-singlet non-relativistic QCD (NRQCD) framework , while the resonance contribution is described through effective $X\gamma^\ast\gamma$ and $X_{HH}$ vertices.

We investigate how channel-dependent interference between the tensor \(X(6900)\) signal and the nonresonant continuum background modifies the \(M_{HH}\) line shape. Comparing the $J/\psi J/\psi\gamma$ and $\eta_c\eta_c\gamma$ channels is particularly useful because the two processes involve different continuum amplitudes and effective $X_{HH}$ vertices. Their resonance strengths and interference patterns may therefore differ between the vector--vector and pseudoscalar--pseudoscalar final-state configurations. For each final‑state channel, we analyze the corresponding invariant-mass distributions and integrated cross sections for different relative phases and $X\gamma^\ast\gamma$ couplings.

The remainder of this paper is organized as follows.  In Sec.~II, we formulate the continuum--resonance framework. In Sec.~III, we present the kinematics and phase-space formulae for $e^+e^-\to H H\gamma$. Numerical results for the $M_{HH}$ distributions and integrated cross sections are given in Sec.~IV, with emphasis on the interference pattern for distinct final‑state configurations, as well as its dependence on the relative phase and \(X\gamma^*\gamma\) coupling. Section~V contains a brief summary.

\section{Continuum--resonance framework}

\begin{figure}[t]
	\centering
	\includegraphics[width=\textwidth]{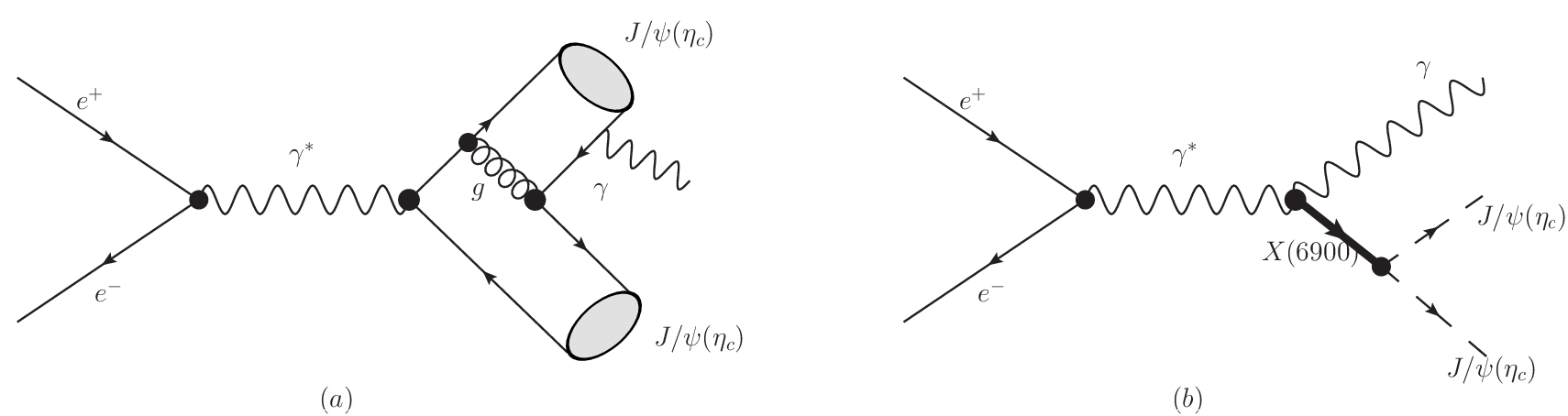}
	\caption{
		Representative diagrams for $e^+e^- \to \gamma^\ast\to HH\gamma$. Panel (a) shows the nonresonant color-singlet NRQCD contribution, while panel (b) represents the resonance mechanism $e^+e^-\to \gamma^\ast\to\gamma X(6900)\to\gamma HH$.}
	\label{fig:feynman}
\end{figure}

The  amplitude is written as the coherent sum
\begin{equation}
	\mathcal A_{\rm full}^{H}
	=
	\mathcal A_{\rm cont}^{H}
	+
	e^{i\phi_X}\mathcal A_{\rm res}^{H},
	\label{eq:Mfull}
\end{equation}
where $\mathcal A_{\rm cont}^{H}$ is the nonresonant continuum amplitude, $\mathcal A_{\rm res}^{H}$ describes the intermediate $X(6900)$ contribution, and $\phi_X$ is their unknown relative phase. The same convention is used for $H=J/\psi$ and $H=\eta_c$. Representative diagrams for the continuum and resonance mechanisms are shown in Fig.~\ref{fig:feynman}.

\subsection{Nonresonant continuum}

Within the NRQCD framework, the differential cross section for the continuum contribution is written as
\begin{equation}
	\frac{d\sigma_{\rm cont}^{H}}{dM_{HH}}
	=
	\frac{d\hat{\sigma}_{\rm cont}^{H}}{dM_{HH}}
	\left\langle 0
	\left|\mathcal O_{n_H}^{H}\right|
	0\right\rangle^2 .
	\label{eq:dsigma_cont_R0}
\end{equation}
where $d\hat{\sigma}_{\rm cont}^{H}/dM_{HH}$ is the short-distance coefficient. The color-singlet long-distance matrix element is related to the radial wave function at the origin by~\cite{Eichten:1995ch}
\begin{equation}
	\left\langle 0
	\left|\mathcal O_{n_H}^{H}\right|
	0\right\rangle
	=
	2N_c(2J_H+1)\frac{|R_S(0)|^2}{4\pi},
	\label{eq:LDME_RS}
\end{equation}
where $J_H$ denotes the spin of the corresponding charmonium.

\subsection{Tensor-resonance amplitude}

We adopt the $J^{PC}=2^{++}$ assignment as a representative tensor scenario for $X(6900)$. The resonance amplitude is factorized as
\begin{equation}
	\mathcal A_{\rm res}^{H}
	=
	\mathcal A_{\gamma^\ast\to\gamma X}
	BW_X(M_{HH}^2)
	\mathcal A_{X\to HH}
	\label{eq:Mres_factor}
\end{equation}
where a fixed-width Breit--Wigner form is used,
\begin{equation}
	BW_X(M_{HH}^2)
	=
	\frac{i}
	{M_{HH}^2-M_X^2+iM_X\Gamma_X}.
	\label{eq:BW}
\end{equation}
The relative phase $\phi_X$ in Eq.~\eqref{eq:Mfull} is defined with respect to this propagator convention.

Let
\begin{equation}
	P=p_3+p_4,
	\qquad
	P^2=M_{HH}^2 .
\end{equation}
The sum over the five spin‑2 polarization states is carried out using the covariant projector~\cite{Zemach:1965,Filippini:1995,Chung:1997qd}:
\begin{equation}
	\sum_{\lambda_X}
	\epsilon_{X,\alpha\beta}(P,\lambda_X)
	\epsilon_{X,\rho\sigma}^{\ast}(P,\lambda_X)
	=
	\Pi_{\alpha\beta,\rho\sigma}(P),
	\label{eq:spin2_pol_sum}
\end{equation}
where
\begin{equation}
	\Pi_{\alpha\beta,\rho\sigma}(P)
	=
	\frac{1}{2}
	\left(
	\bar g_{\alpha\rho}\bar g_{\beta\sigma}
	+
	\bar g_{\alpha\sigma}\bar g_{\beta\rho}
	\right)
	-
	\frac{1}{3}
	\bar g_{\alpha\beta}\bar g_{\rho\sigma},
	\qquad
	\bar g_{\mu\nu}
	=
	-g_{\mu\nu}
	+
	\frac{P_\mu P_\nu}{P^2}.
	\label{eq:spin2proj}
\end{equation}

For the benchmark analysis, we retain the following minimal effective
Lorentz structures:
\begin{align}
	\mathcal A_{X\to J/\psi J/\psi}
	&=
	g_{T\psi}
	\epsilon_X^{\mu\nu}
	\varepsilon_{3\mu}^{\ast}
	\varepsilon_{4\nu}^{\ast},
	\label{eq:TXpsipsi_vertex}
	\\
	\mathcal A_{X\to\eta_c\eta_c}
	&=
	g_{T\eta}
	\epsilon_X^{\mu\nu}r_\mu r_\nu,
	\qquad
	r=p_3-p_4 .
	\label{eq:TXetaeta_vertex}
\end{align}
Here $\varepsilon_3$ and $\varepsilon_4$ are the polarization vectors of the two final-state $J/\psi$ mesons. Additional independent tensor structures are not included in the present benchmark model.

The on-shell electromagnetic vertex is taken as
\begin{equation}
	\mathcal A_{X\to\gamma\gamma}
	=
	g_{T\gamma}
	\epsilon_X^{\mu\nu}
	\left(
	F_{1,\mu}{}^{\alpha}F_{2,\nu\alpha}
	+
	F_{2,\mu}{}^{\alpha}F_{1,\nu\alpha}
	\right),
	\label{eq:TXgammagamma_vertex}
\end{equation}
where
\begin{equation}
	F_a^{\mu\nu}
	=
	q_a^\mu\varepsilon_a^\nu
	-
	q_a^\nu\varepsilon_a^\mu,
	\qquad
	a=1,2 .
\end{equation}
Here, \(F_a^{\mu\nu}\) denotes the electromagnetic field-strength tensor associated with the \(a\)-th photon, while \(q_a\) and \(\varepsilon_a\) are its four-momentum and polarization vector, respectively.
The two terms in Eq.~\eqref{eq:TXgammagamma_vertex} make the Bose symmetry under photon exchange explicit.
In the production amplitude, one photon momentum is continued to the time-like value $k$, while the other is identified with the final-state photon of momentum $p_5$. Possible additional structures associated with the off-shell photon are not considered.

\subsection{Effective couplings and off-shell form factor}

For a spin-2 particle decaying into two identical particles,
\begin{equation}
	\Gamma(X\to ab)
	=
	\frac{|\bm p|}{16\pi M_X^2}
	\overline{
		\left|\mathcal A_{X\to ab}\right|^2},
	\qquad
	\overline{|\mathcal A|^2}
	=
	\frac{1}{5}
	\sum_{\lambda_X}|\mathcal A|^2 .
	\label{eq:general_two_body_width}
\end{equation}
The factor $1/(2J_X+1)=1/5$ 
accounts for averaging over the five spin‑2 polarization states of the resonance when evaluating an unpolarized decay width..

Using the effective vertices above, the relevant partial widths are
\begin{align}
	\Gamma(X\to\gamma\gamma)
	&=
	\frac{g_{T\gamma}^2 M_X^3}{80\pi},
	\label{eq:Gamma_gamgam}
	\\
	\Gamma(X\to J/\psi J/\psi)
	&=
	\frac{g_{T\psi}^2 |\bm p_\psi|}{16\pi M_X^2}
\frac{2r_\psi^2+6r_\psi+7}{15},
	\label{eq:Gamma_psipsi}
	\\
	\Gamma(X\to\eta_c\eta_c)
	&=
	\frac{2g_{T\eta}^2|\bm p_\eta|^5}
	{15\pi M_X^2}.
	\label{eq:Gamma_etaeta}
\end{align}
where
\begin{equation}
	|\bm p_\psi|
	=
	\sqrt{\frac{M_X^2}{4}-m_{J/\psi}^2},
	\qquad
	|\bm p_\eta|
	=
	\sqrt{\frac{M_X^2}{4}-m_{\eta_c}^2},
	\qquad
	r_\psi
	=
	\frac{M_X^2}{4m_{J/\psi}^2}.
	\label{eq:hadronic_momenta}
\end{equation}

Using the hadronic inputs listed in Table~\ref{tab:inputs}, we obtain
\begin{equation}
	\begin{aligned}
		g_{T\psi} &= 10.510~{\rm GeV},
		&
		g_{T\eta} &= 0.0745~{\rm GeV}^{-1},
		\\
		g_{T\gamma}^{(\mathrm{I})}
		&= 2.41\times10^{-4}~{\rm GeV}^{-1},
		&
		g_{T\gamma}^{(\mathrm{II})}
		&= 2.58\times10^{-4}~{\rm GeV}^{-1}.
	\end{aligned}
	\label{eq:effective_couplings}
\end{equation}

The diphoton width fixes only the on-shell normalization of the electromagnetic vertex. In the present process, the initial photon is time-like, with $k^2=s>0$, whereas the final-state photon is on shell.
Since the timelike $X\gamma^\ast\gamma$ transition form factor is not presently known, we adopt the following phenomenological parametrization:
\begin{equation}
	g_{T\gamma^\ast\gamma}^{(i)}(s)
	=
	F_{\exp}(s)g_{T\gamma}^{(i)}
	\qquad
	i={\rm I,II},
	\label{eq:gTgamma_offshell}
\end{equation}
with
\begin{equation}
	F_{\exp}(s)
	=
	\exp\left(-\frac{s}{\Lambda_X^2}\right).
	\label{eq:Fexp}
\end{equation}
Motivated by exponential form factors used in related phenomenological studies~\cite{Goerke:2016hxf,Lu:2025lyu,Dubnicka:2010kz}, we take $\Lambda_X=M_X$ as the benchmark choice. This parametrization satisfies $F_{\exp}(0)=1$ and suppresses the transition amplitude smoothly at large time-like virtuality. At fixed $\sqrt{s}$, $F_{\exp}(s)$ is independent of $M_{HH}$ and therefore modifies the relative normalization of the resonance and interference contributions without introducing an additional $M_{HH}$ dependence.

\subsection{Coherent invariant-mass distribution}

For each final state and electromagnetic input, the full distribution
is decomposed as
\begin{equation}
	\frac{d\sigma_{\rm full}^{H,(i)}}{dM_{HH}}
	=
	B_H(M_{HH})
	+
	R_H^{(i)}(M_{HH})
	+
	I_H^{(i)}(M_{HH},\phi_X),
	\label{eq:full_decompose}
\end{equation}
where $B_H(M_{HH})$, defined by
\begin{equation}
	B_H(M_{HH})
	\equiv
	\frac{d\sigma_{\rm cont}^{H}}{dM_{HH}}
\end{equation}
is the NRQCD continuum contribution from Eq.~\eqref{eq:dsigma_cont_R0},
$R_H^{(i)}$ is obtained from
$\left|\mathcal A_{\rm res}^{H,(i)}\right|^2$, and the interference
term is proportional to
\begin{equation}
	I_H^{(i)}(M_{HH},\phi_X)
	\propto
	2{\rm Re}
	\left[
	e^{i\phi_X}
	\mathcal A_{\rm res}^{H,(i)}
	\left(\mathcal A_{\rm cont}^{H}\right)^\ast
	\right].
	\label{eq:interference}
\end{equation}
The resonance contribution is quadratic in
$g_{T\gamma}^{(i)}F_{\exp}(s)$, whereas the interference term is linear in this quantity. This difference will be used below to assess the dependence of the $M_{HH}$ line shape on the electromagnetic input.

\section{Kinematics and phase space}

With the momentum assignments defined in Eq.~\eqref{eq:all}, the relevant kinematic relations are
\begin{equation}
	k=l_1+l_2,\qquad
	k^2=s,\qquad
	p_5^2=0,\qquad
	p_3^2=p_4^2=m_H^2 .
\end{equation}
The kinematic range of $M_{HH}$ in Eq.~\eqref{eq:dsigma_cont_R0} is
\begin{equation}
	2m_H\leq M_{HH}\leq \sqrt{s}.
	\label{eq:M}
\end{equation}

In the $e^+e^-$ center-of-mass frame,
$k=(\sqrt{s},\mathbf{0})$, and energy conservation gives
\begin{equation}
	E_5=\frac{s-M_{HH}^2}{2\sqrt{s}},
	\qquad
	E_3+E_4=\frac{s+M_{HH}^2}{2\sqrt{s}} .
	\label{eq:energy}
\end{equation}
At fixed $M_{HH}$, the allowed range of $E_3$ is
\begin{equation}
	E_3^{\min,\max}(M_{HH})
	=
	\frac{s+M_{HH}^2}{4\sqrt{s}}
	\mp
	\frac{s-M_{HH}^2}{4\sqrt{s}M_{HH}}
	\sqrt{M_{HH}^2-4m_H^2}.
	\label{eq:E3limits}
\end{equation}
The same limits apply to the continuum, resonance, and interference
contributions.

The invariant-mass differential cross section is then given by
\begin{equation}
	\frac{d\sigma}{dM_{HH}}
	=
	\frac{1}{2!}
	\frac{1}{4}
	\frac{1}{s^2}
	\left(\frac{4s}{3}e^2\right)
	\frac{1}{2s}
	\frac{1}{4(2\pi)^3}
	\frac{M_{HH}}{\sqrt{s}}
	\int_{E_3^{\min}}^{E_3^{\max}}
	dE_3
	\left|\mathcal A_{\rm red}\right|^2 ,
	\label{eq:dsigmadM}
\end{equation}
where the factor $1/4$ averages over the initial lepton spins, while
$1/2!$ accounts for the two identical charmonia in the final state.
Here, $\mathcal A_{\rm red}$ denotes the amplitude for the reduced subprocess
$\gamma^\ast\to HH\gamma$, with the leptonic part factored out.

\section{Numerical results and discussion}

\subsection{Input parameters}
The input parameters used in the numerical analysis are summarized in Table~\ref{tab:inputs}. Except for the two-photon width $\Gamma_{\gamma\gamma}$, which distinguishes Model I and Model II, all other parameters are fixed throughout the calculation.

\begin{table}[t]
	\caption{ Input parameters for the tensor-resonance calculation. Model I and Model II correspond to two representative inputs for $\Gamma_{\gamma\gamma}$ used to determine $g_{T\gamma}$~\cite{Sang:2023ncm,Lu:2025lyu,ParticleDataGroup:2024cfk}.}
	\label{tab:inputs}
	\begin{ruledtabular}
		\begin{tabular}{lc}
			Quantity & Value \\
			$\sqrt{s}$ & $10.58~{\rm GeV}$ \\
			$M_X$ & $6.905~{\rm GeV}$ \\
			$\Gamma_X$ & $0.168~{\rm GeV}$ \\
			$m_c$ & $1.5~{\rm GeV}$ \\
			$m_{J/\psi}$ & $3.0969~{\rm GeV}$ \\
			$m_{\eta_c}$ & $2.984~{\rm GeV}$ \\
			$\Gamma_{\psi\psi}$ & $82~{\rm MeV}$ \\
			$\Gamma_{\eta\eta}$ & $77~{\rm keV}$ \\
			$\Gamma_{\gamma\gamma}^{\rm I}$ & $7.6\times10^{-8}~{\rm GeV}$ \\
			$\Gamma_{\gamma\gamma}^{\rm II}$ & $8.7\times10^{-8}~{\rm GeV}$ \\
		\end{tabular}
	\end{ruledtabular}
\end{table}

\subsection{Invariant-mass distributions and phase dependence}

\begin{figure*}[t]
	\centering
	\begin{minipage}{0.48\textwidth}
		\centering
		\includegraphics[width=\textwidth]{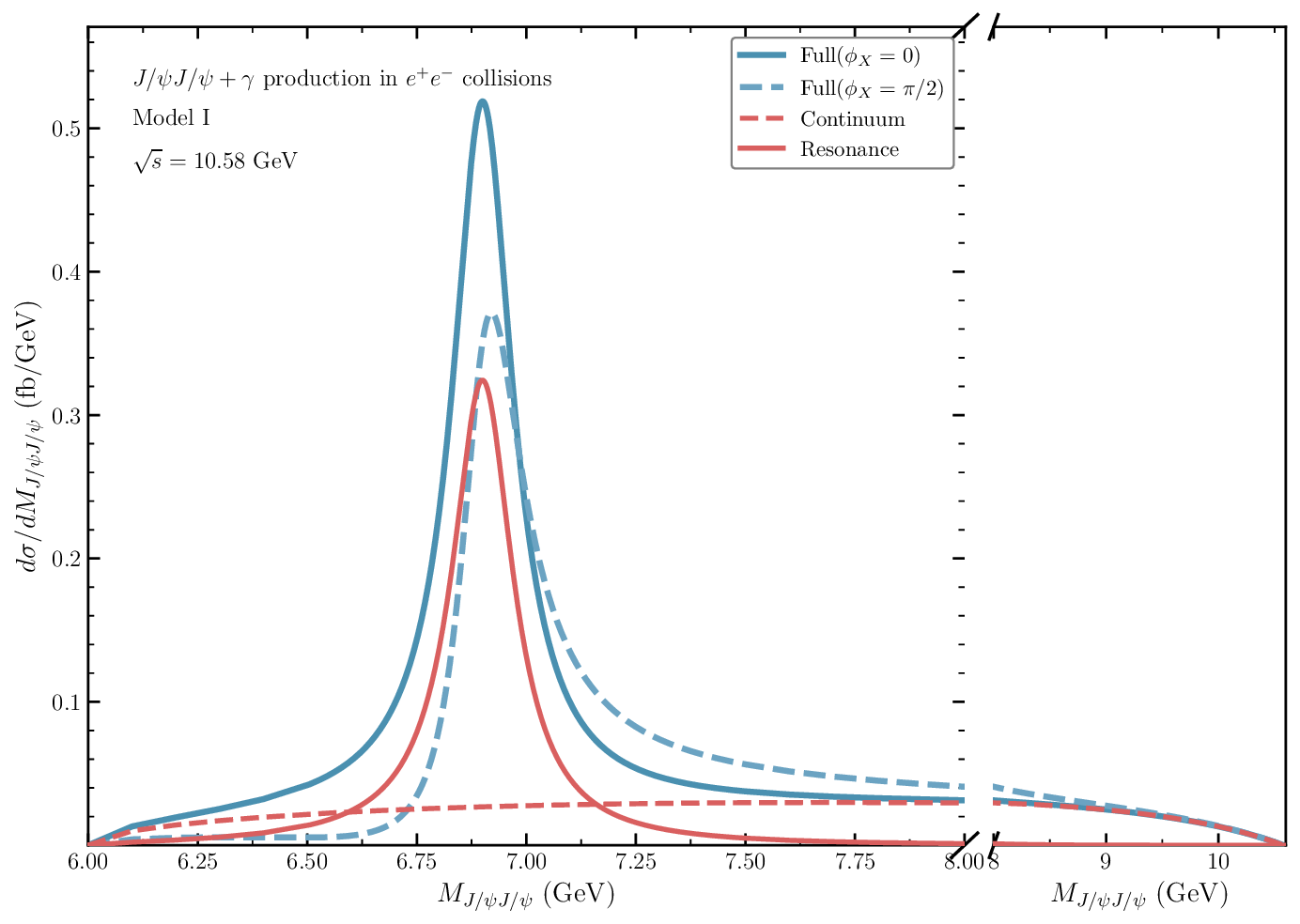}
		\vspace{-2mm}
		\centerline{(a) $J/\psi J/\psi\gamma$, Model I}
	\end{minipage}
	\hfill
	\begin{minipage}{0.48\textwidth}
		\centering
		\includegraphics[width=\textwidth]{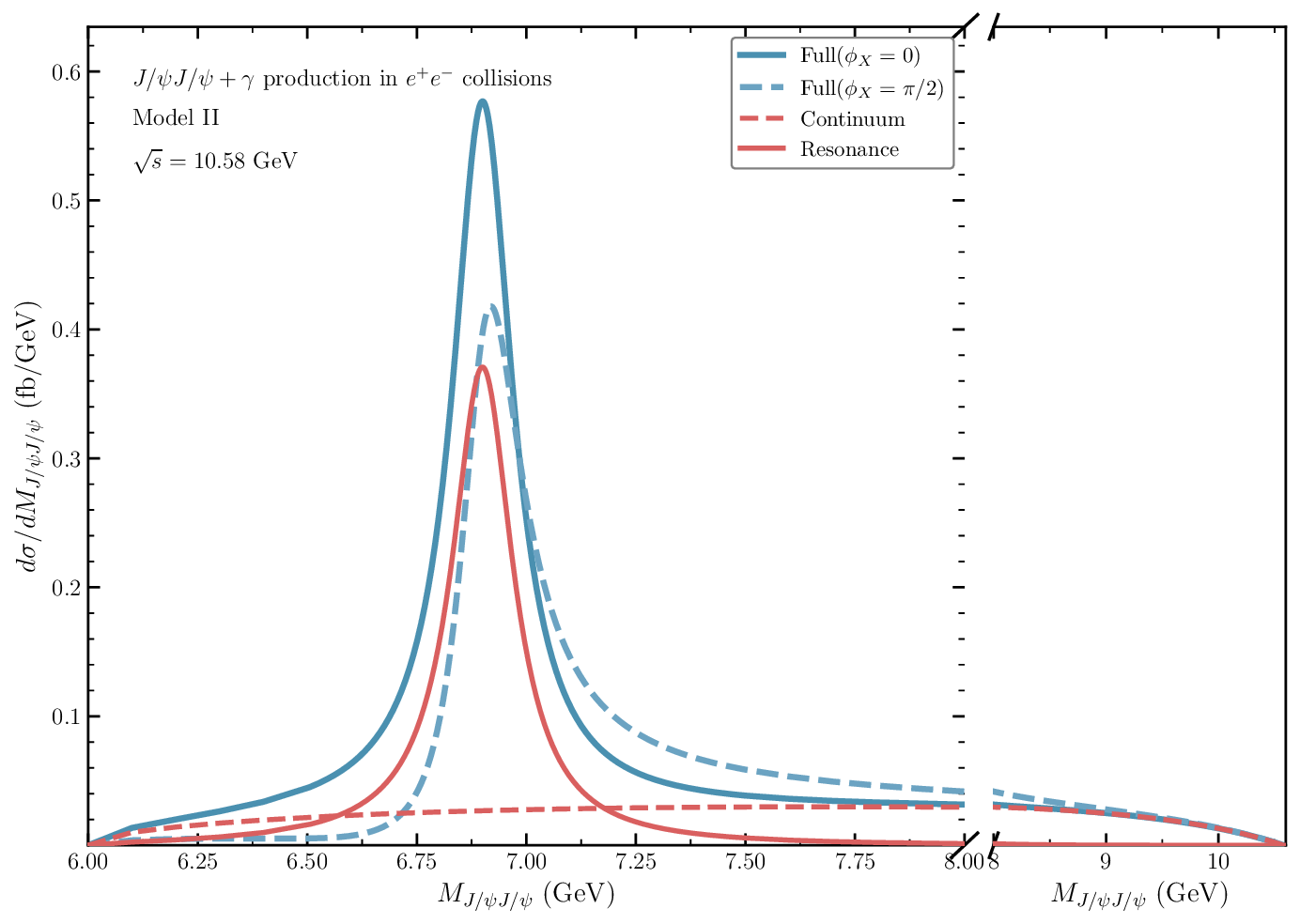}
		\vspace{-2mm}
		\centerline{(b) $J/\psi J/\psi\gamma$, Model II}
	\end{minipage}

	\vspace{3mm}
	
	\begin{minipage}{0.48\textwidth}
		\centering
		\includegraphics[width=\textwidth]{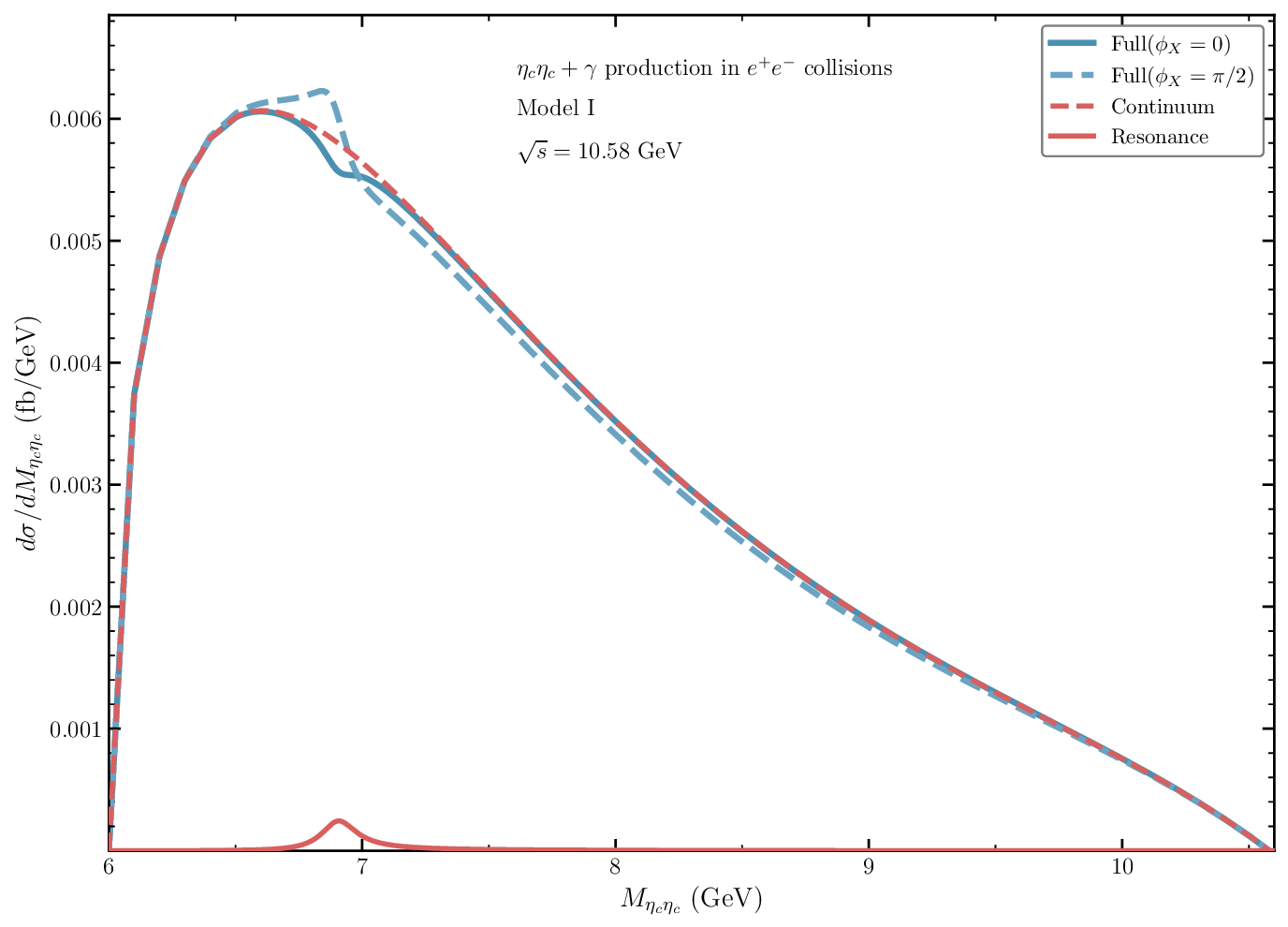}
		\vspace{-2mm}
		\centerline{(c) $\eta_c\eta_c\gamma$, Model I}
	\end{minipage}
	\hfill
	\begin{minipage}{0.48\textwidth}
		\centering
		\includegraphics[width=\textwidth]{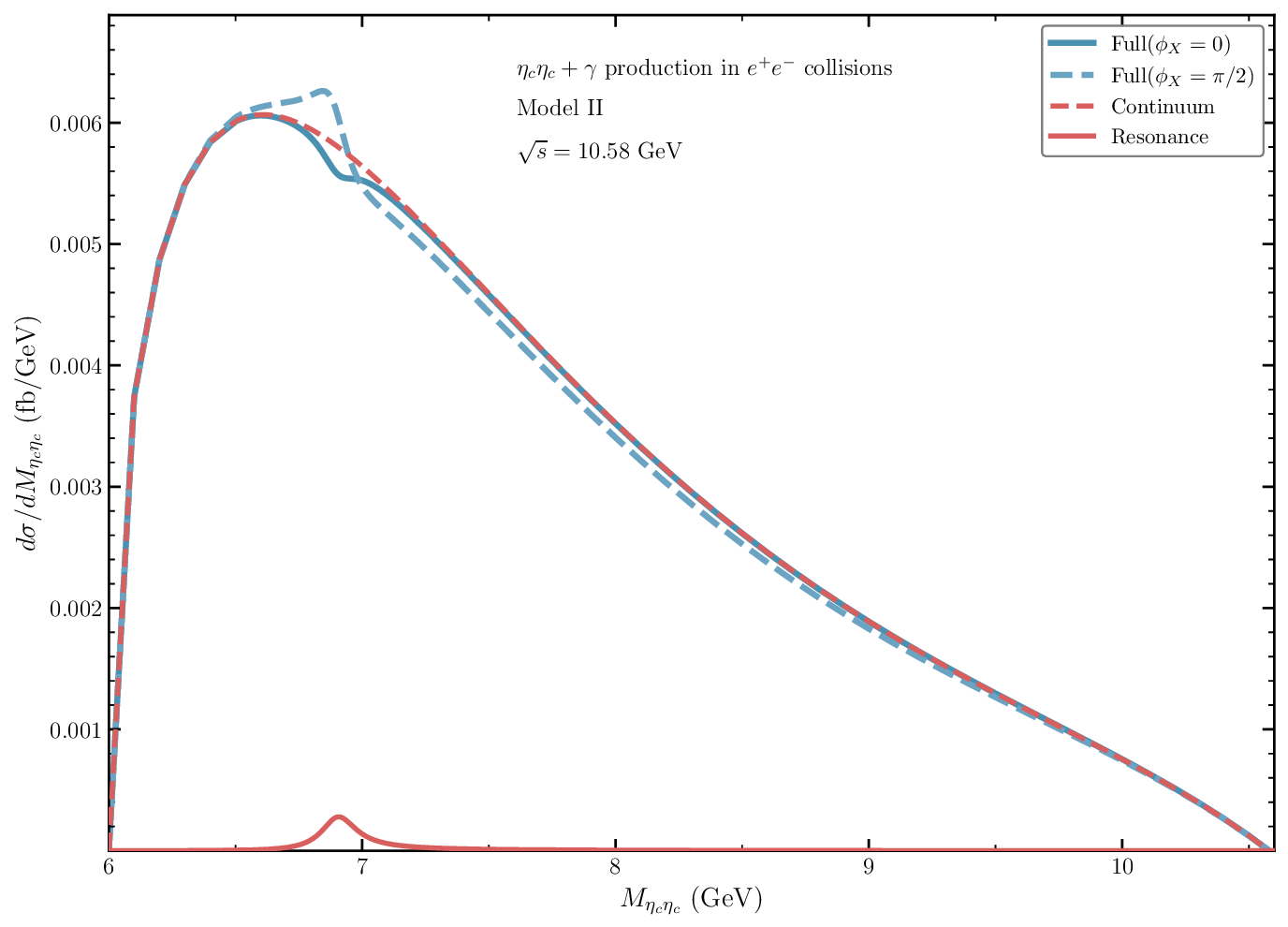}
		\vspace{-2mm}
		\centerline{(d) $\eta_c\eta_c\gamma$, Model II}
	\end{minipage}
	
	\caption{Invariant-mass distributions $d\sigma/dM_{HH}$ for
		$e^+e^-\to HH\gamma$ at $\sqrt{s}=10.58~{\rm GeV}$.
		Panels (a) and (b) correspond to $J/\psi J/\psi\gamma$ in Models I and II, respectively, while panels (c) and (d) show the corresponding
		$\eta_c\eta_c\gamma$ results. The continuum-only and
		resonance-only contributions are displayed together with
		the coherent distributions for $\phi_X=0$ and $\pi/2$.}
	\label{fig:all}

\end{figure*}

Figure~\ref{fig:all} compares the distributions for the nonresonant continuum, resonance‑only contribution, and full coherently‑summed result with interference for $e^+e^-\to HH\gamma$ at $\sqrt{s}=10.58~{\rm GeV}$. For illustration, the coherent results are shown for the two representative phases $\phi_X=0$ and $\pi/2$.

For $J/\psi J/\psi\gamma$, the resonance-only contribution produces a pronounced localized structure near the nominal resonance mass $M_X$, whereas the continuum varies smoothly and remains comparatively small. Despite its smaller magnitude, the continuum amplitude gives rise to a
substantial interference contribution. Within the phase and coupling conventions adopted here, the coherent distribution for $\phi_X=0$ lies above the resonance-only result in the peak region, corresponding to constructive interference. For $\phi_X=\pi/2$, the interference redistributes strength across the resonance region and produces a more asymmetric line shape. Model II increases the resonance normalization without changing these qualitative features.

The $\eta_c\eta_c\gamma$ channel exhibits a different hierarchy. Its continuum distribution has a broad maximum below $M_X$ and decreases smoothly through the resonance region. The resonance-only contribution is much smaller and is concentrated near $M_{\eta_c\eta_c}\simeq M_X$. The tensor contribution therefore does not appear as an isolated dominant peak. Depending on the relative phase, interference instead produces a localized suppression or enhancement near the nominal resonance mass. The close agreement between Models I and II over most of the invariant-mass range further shows that the full distribution in this channel is controlled mainly by the continuum.

This behavior follows from the different dependence of the two resonance-related terms on the resonance amplitude. The interference term is linear in $\mathcal A_{\rm res}$, whereas the resonance-squared term is quadratic in $\mathcal A_{\rm res}$. A relatively small resonance amplitude may therefore produce a non-negligible local modification through interference even when the resonance-only contribution is strongly suppressed.

\begin{figure*}[t]
	\centering
	\begin{minipage}{0.48\textwidth}
		\centering
		\includegraphics[width=\textwidth]{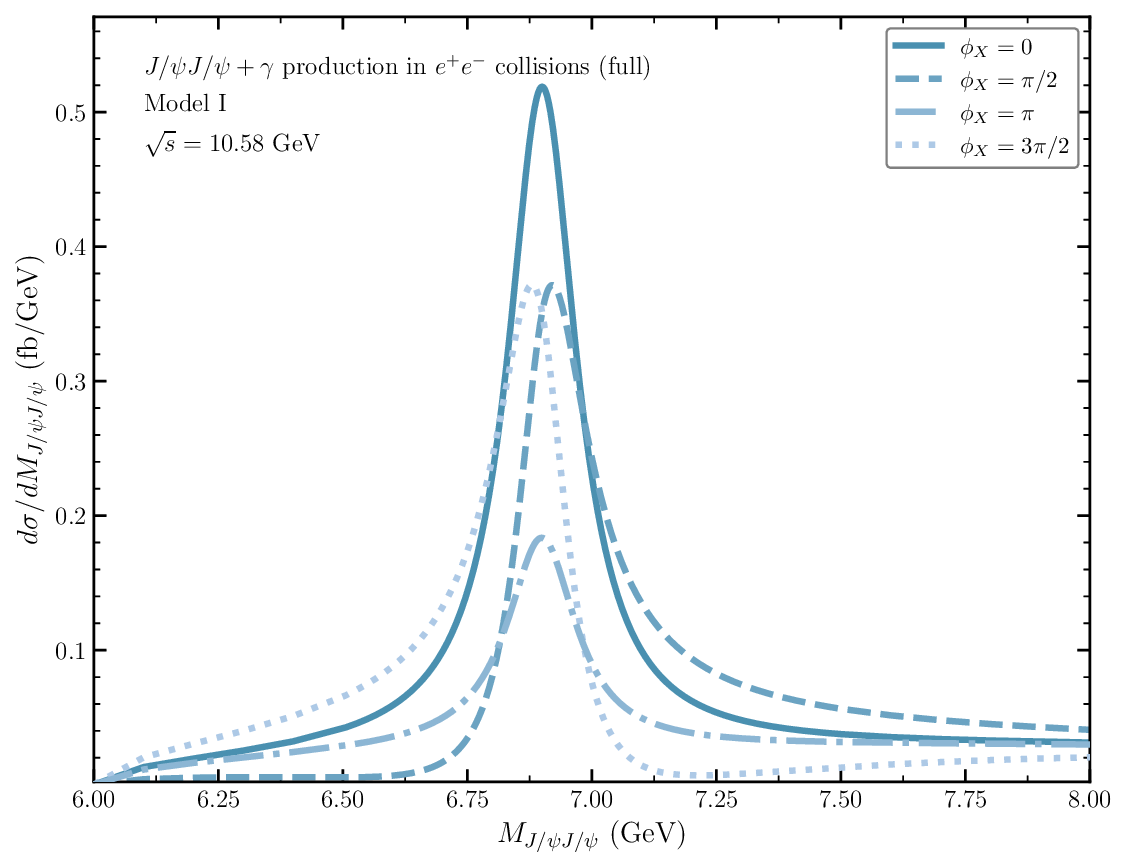}
		\vspace{-2mm}
		\centerline{(a) $J/\psi J/\psi\gamma$, Model I}
	\end{minipage}
	\hfill
	\begin{minipage}{0.48\textwidth}
		\centering
		\includegraphics[width=\textwidth]{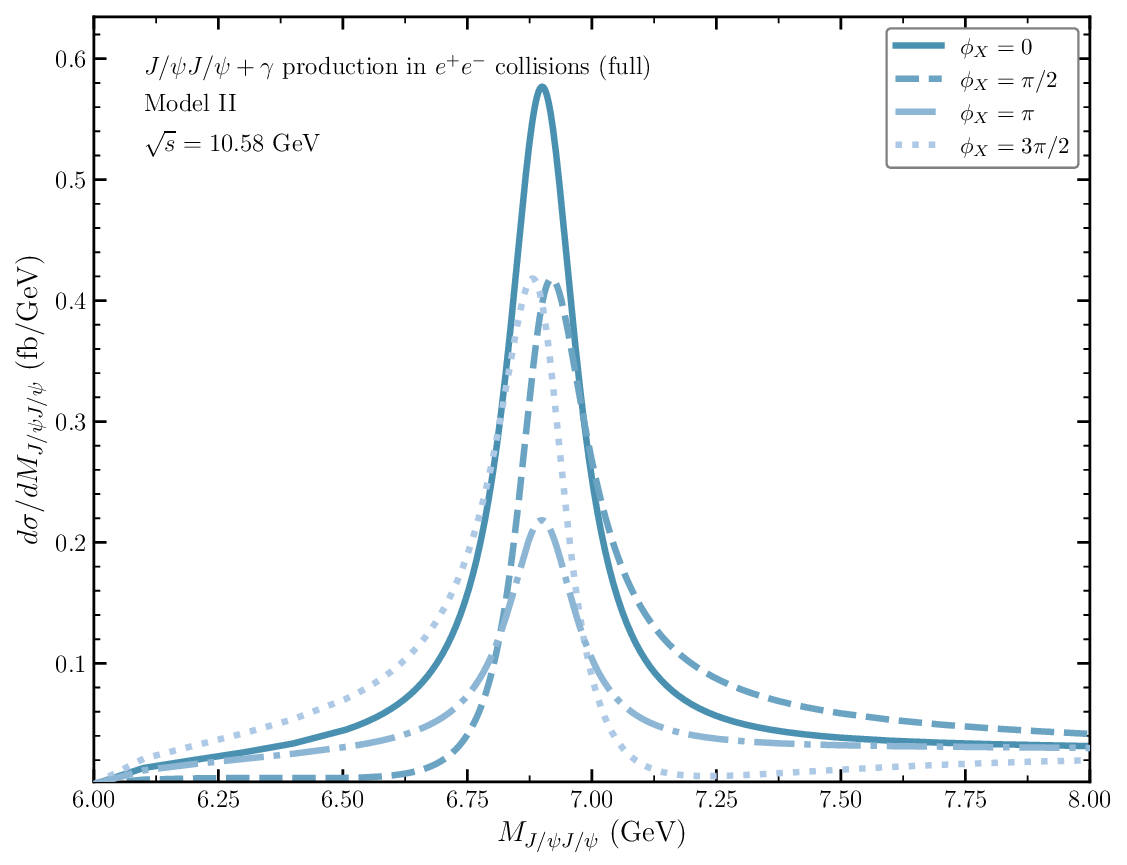}
		\vspace{-2mm}
		\centerline{(b) $J/\psi J/\psi\gamma$, Model II}
	\end{minipage}

	\vspace{3mm}
	
	\begin{minipage}{0.48\textwidth}
		\centering
		\includegraphics[width=\textwidth]{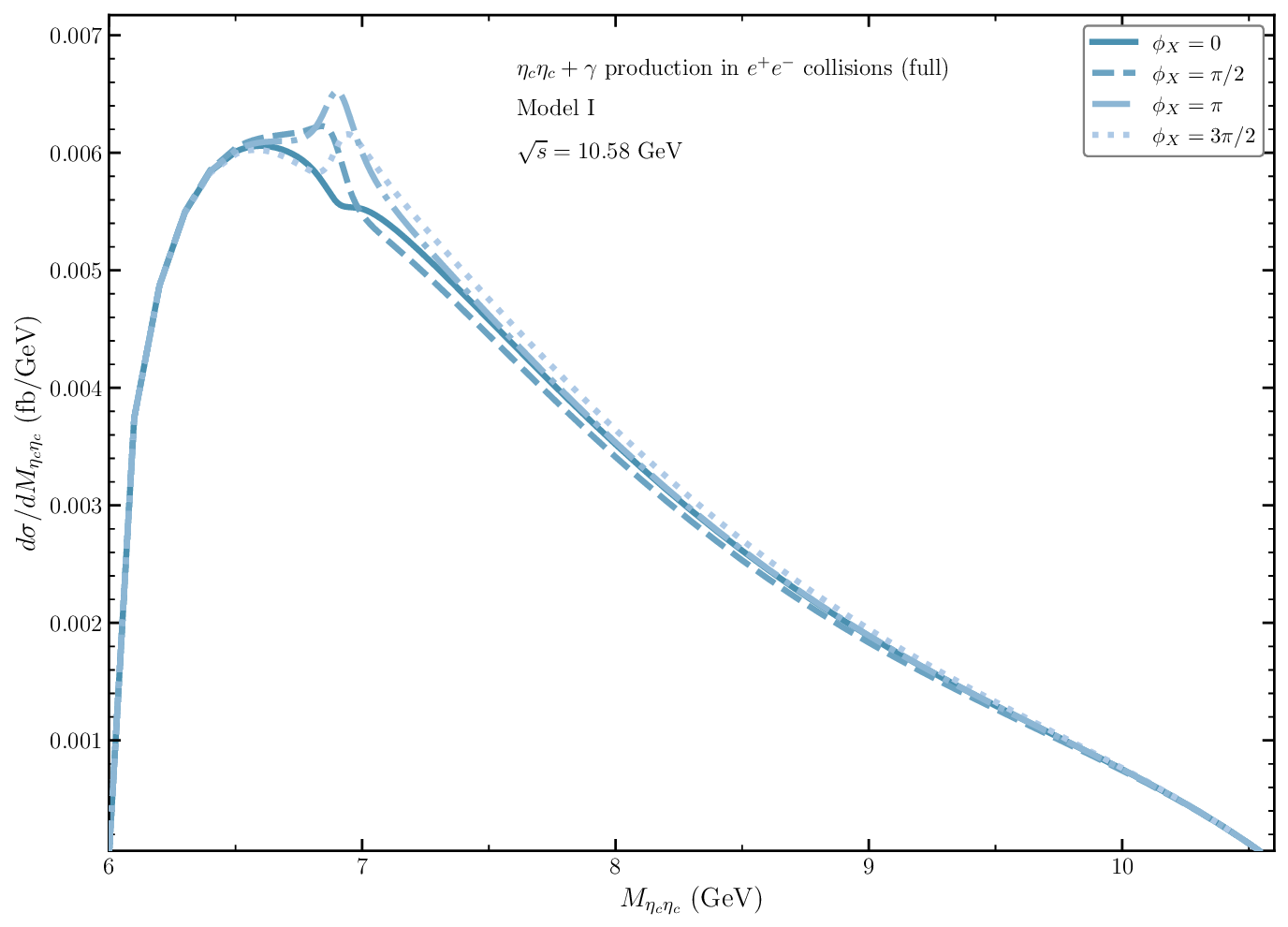}
		\vspace{-2mm}
		\centerline{(c) $\eta_c\eta_c\gamma$, Model I}
	\end{minipage}
	\hfill
	\begin{minipage}{0.48\textwidth}
		\centering
		\includegraphics[width=\textwidth]{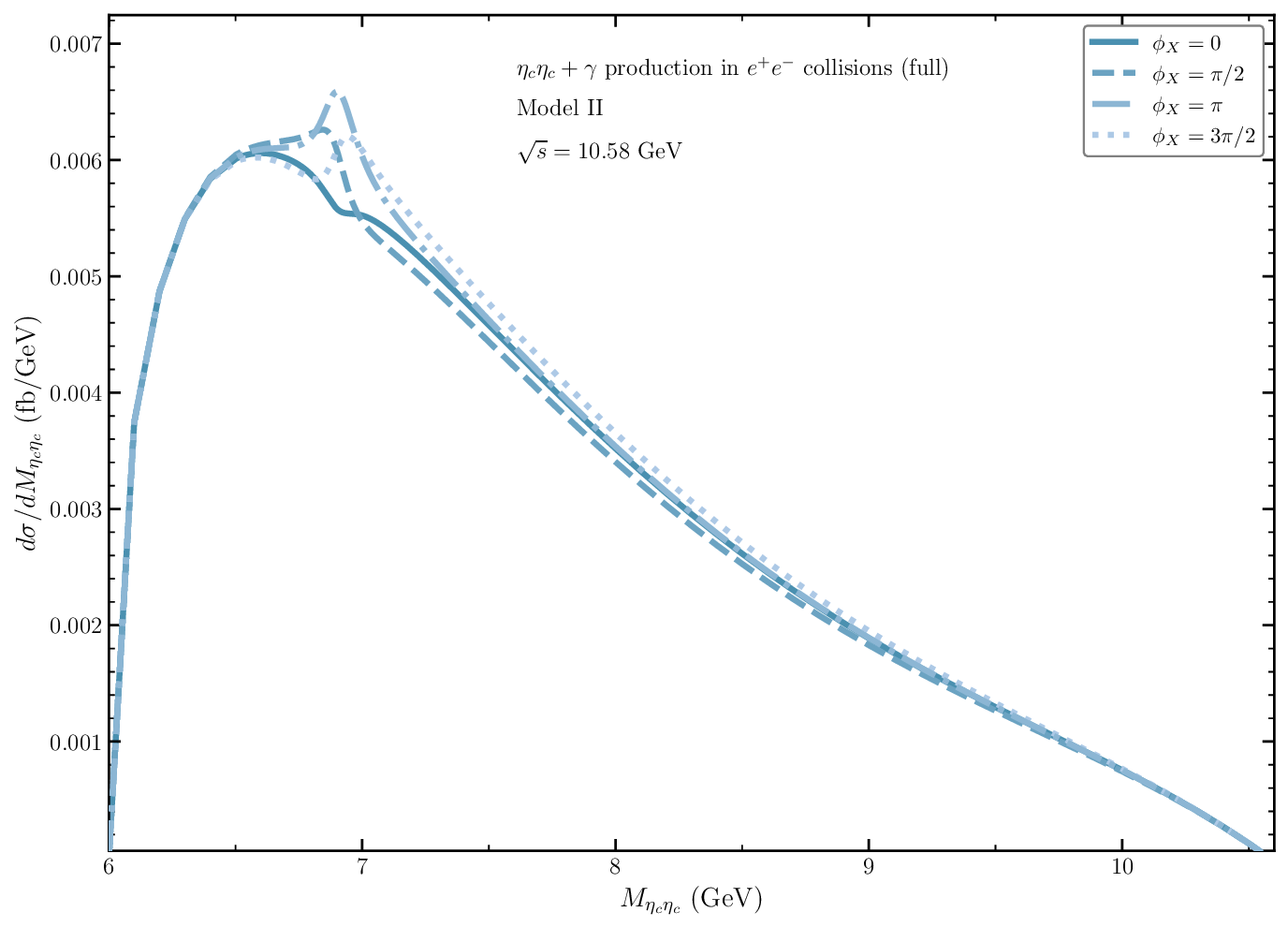}
		\vspace{-2mm}
		\centerline{(d) $\eta_c\eta_c\gamma$, Model II}
	\end{minipage}
	
	\caption{
		Coherent invariant-mass distributions for
		$e^+e^-\to HH\gamma$ at $\sqrt{s}=10.58~{\rm GeV}$
		for $\phi_X=0$, $\pi/2$, $\pi$, and $3\pi/2$.
		The phase dependence is concentrated near
		$M_{HH}\simeq M_X$, where the resonance contribution is
		largest. The two quadrature phases produce asymmetric
		distortions with opposite orientations across the resonance
		region.
	}
	\label{fig:phase}
	
\end{figure*}

The phase dependence is examined in more detail in Fig.~\ref{fig:phase}. Since the relative phase is not determined within the present effective framework, we consider
\begin{equation}
	\phi_X
	=
	0,\quad
	\frac{\pi}{2},\quad
	\pi,\quad
	\frac{3\pi}{2}.
\end{equation}
These values are used to sample representative interference patterns rather than to determine the physical relative phase. The same benchmark values are applied separately to the two channels for ease of comparison; this does not imply that their physical relative phases must be identical.

With the Breit--Wigner convention adopted in this work, the
interference contribution can be written schematically as
\begin{equation}
	\frac{d\sigma_{\rm int}^{H}}{dM_{HH}}
	=
	C_H(M_{HH})\cos\phi_X
	+
	S_H(M_{HH})\sin\phi_X,
	\label{eq:phase_decomposition}
\end{equation}
where $C_H(M_{HH})$ and $S_H(M_{HH})$ are real functions after phase-space integration. If the continuum amplitude varies slowly and is approximately real across the resonance region, the term proportional to $\cos\phi_X$ produces an approximately symmetric enhancement or suppression near $M_{HH}\simeq M_X$, whereas the term proportional to $\sin\phi_X$ changes sign across the resonance and generates an asymmetric peak--dip or dip--peak profile. Interference has also been found to be important in phenomenological descriptions of the di-$J/\psi$ spectrum at hadron colliders
~\cite{CMS:2023owd}, although the underlying production mechanism differs from that considered here.

Within the sign convention adopted for the effective couplings, $\phi_X=0$ gives the largest enhancement in the $J/\psi J/\psi\gamma$ channel, whereas $\phi_X=\pi$ gives the most strongly suppressed resonant structure. The phases $\pi/2$ and $3\pi/2$ redistribute strength toward opposite sides of $M_X$ and produce complementary asymmetric profiles. The unknown phase can thus affect both the apparent peak height and the local line shape.

For $\eta_c\eta_c\gamma$, the phase ordering near the resonance is different within the same convention. The choice $\phi_X=0$ gives a localized suppression, whereas $\phi_X=\pi$ produces a localized enhancement. The two quadrature phases again generate asymmetric distortions with opposite orientations. The same numerical value of $\phi_X$ therefore need not correspond to the same constructive or destructive pattern in the two channels.

This cross-channel comparison is convention dependent. The partial widths determine only the magnitudes of the hadronic couplings $g_{T\psi}$ and $g_{T\eta}$, not their relative signs. Reversing the sign of one hadronic coupling is equivalent to shifting the corresponding phase by $\pi$. The phase ordering between the two channels should therefore be understood within the coupling convention used in the present calculation.

The different line shapes nevertheless reflect genuine differences in the two amplitudes. As shown in Eqs.~\eqref{eq:TXpsipsi_vertex} and \eqref{eq:TXetaeta_vertex}, the $J/\psi J/\psi$ vertex is controlled by the vector polarizations, whereas the $\eta_c\eta_c$ vertex contains two powers of the relative momentum. These structures contract differently with the corresponding NRQCD continuum amplitudes. Together with the smaller input partial width for $X\to\eta_c\eta_c$, this accounts for the weaker resonance contribution in the $\eta_c\eta_c\gamma$ channel.

The distributions shown in Figs.~\ref{fig:all} and \ref{fig:phase} are theory-level predictions without detector smearing, reconstruction efficiencies, or experimental backgrounds. The corresponding structures in reconstructed distributions may therefore be less pronounced.

\subsection{Model dependence and integrated cross sections}

Models I and II differ through the electromagnetic transition coupling $g_{T\gamma}^{(i)}$, while the same off-shell form factor is used in both cases. For fixed hadronic couplings and continuum amplitudes, the resonance-squared and interference terms scale as
\begin{equation}
	R_H^{(i)}(M_{HH})
	\propto
	\left[
	g_{T\gamma}^{(i)}F_{\exp}(s)
	\right]^2,
	\qquad
	I_H^{(i)}(M_{HH},\phi_X)
	\propto
	g_{T\gamma}^{(i)}F_{\exp}(s).
	\label{eq:model_scaling}
\end{equation}
The resonance-squared term is therefore more sensitive to the
electromagnetic input than the interference term. This difference is most apparent in the $J/\psi J/\psi\gamma$ channel, where the resonance contribution is sizable. In the continuum-dominated $\eta_c\eta_c\gamma$ channel, changing the electromagnetic input has only a small effect on the full distribution.

To complement the differential results, we define cross sections integrated over the full kinematic region and over a symmetric mass window centered on the nominal resonance mass:
\begin{equation}
	\sigma_{\rm full}^{H,(i)}(\phi_X)
	=
	\int_{2m_H}^{\sqrt{s}}
	dM_{HH},
	\frac{d\sigma_{\rm full}^{H,(i)}}{dM_{HH}},
	\label{eq:sigma_full_integrated}
\end{equation}
and
\begin{equation}
	\sigma_{\rm near}^{H,(i)}(\phi_X)
	=
	\int_{M_X-\Delta M}^{M_X+\Delta M}
	dM_{HH},
	\frac{d\sigma_{\rm full}^{H,(i)}}{dM_{HH}}.
	\label{eq:sigma_near_integrated}
\end{equation}

For definiteness, we take $\Delta M=\Gamma_X$, corresponding to one natural width on each side of the nominal mass and hence to a total window of $2\Gamma_X$. The mass intervals used in experimental
analyses of the inclusive di-$J/\psi$ spectrum are generally broader and may contain threshold enhancements, neighboring structures, and nonresonant contributions ~\cite{LHCb:2020bwg,ATLAS:2023bft,CMS:2023owd}. Using the natural width as the characteristic scale provides a uniform benchmark for
quantifying the local resonance and interference contributions
~\cite{Pappadopulo:2014qza}. This choice does not define a unique physical window, and the numerical value of $\sigma_{\rm near}$ should be understood as being tied to this benchmark interval. The full-region cross section characterizes the overall production rate,
whereas $\sigma_{\rm near}$ is more sensitive to the local behavior near $M_X$.

\begin{table}[t]
	\caption{
		Integrated cross sections for the coherent distributions.
		The full-region and near-resonance cross sections are defined
		in Eqs.~\eqref{eq:sigma_full_integrated} and
		\eqref{eq:sigma_near_integrated}, respectively, with
		$\Delta M=\Gamma_X$.
	}
	\label{tab:integrated_cross_sections}
	\begin{ruledtabular}
		\begin{tabular}{lccc}
			Channel and model
			& $\phi_X$
			& $\sigma_{\rm full}$ [fb]
			& $\sigma_{\rm near}$ [fb] \\
			\hline
			$J/\psi J/\psi\gamma$, I
			& $0$ & $0.222$ & $0.100$ \\
			& $\pi/2$ & $0.195$ & $0.070$ \\
			& $\pi$ & $0.139$ & $0.038$ \\
			& $3\pi/2$ & $0.166$ & $0.068$ \\[1mm]
			
			$J/\psi J/\psi\gamma$, II
			& $0$ & $0.236$ & $0.111$ \\
			& $\pi/2$ & $0.207$ & $0.079$ \\
			& $\pi$ & $0.148$ & $0.045$ \\
			& $3\pi/2$ & $0.176$ & $0.077$ \\[1mm]
			
			$\eta_c\eta_c\gamma$, I
			& $0$ & $1.406\times10^{-2}$ & $1.901\times10^{-3}$ \\
			& $\pi/2$ & $1.392\times10^{-2}$ & $1.974\times10^{-3}$ \\
			& $\pi$ & $1.430\times10^{-2}$ & $2.077\times10^{-3}$ \\
			& $3\pi/2$ & $1.445\times10^{-2}$ & $2.004\times10^{-3}$ \\[1mm]
			
			$\eta_c\eta_c\gamma$, II
			& $0$ & $1.406\times10^{-2}$ & $1.901\times10^{-3}$ \\
			& $\pi/2$ & $1.391\times10^{-2}$ & $1.979\times10^{-3}$ \\
			& $\pi$ & $1.430\times10^{-2}$ & $2.090\times10^{-3}$ \\
			& $3\pi/2$ & $1.448\times10^{-2}$ & $2.012\times10^{-3}$ \\
		\end{tabular}
	\end{ruledtabular}
\end{table}
The integrated cross sections are listed in
Table~\ref{tab:integrated_cross_sections}. In the
$J/\psi J/\psi\gamma$ channel, both the full-region and
near-resonance rates depend strongly on $\phi_X$. For Model I,
$\sigma_{\rm near}$ ranges from $0.038$ to $0.100~{\rm fb}$, while Model II gives $0.045$--$0.111~{\rm fb}$. In both models, the largest near-resonance rate occurs at $\phi_X=0$, whereas the smallest is obtained at $\phi_X=\pi$. The variation by more than a factor of two
shows that the yield around $M_X$ remains sensitive to interference even after integration over the symmetric benchmark window.

The two quadrature phases do not generally give identical integrated rates. Although they produce oppositely oriented distortions in the differential distributions, the continuum amplitude, the phase-space weight, and the remaining kinematic factors vary across the physical mass range. The antisymmetric contribution therefore need not vanish after integration.

Model II gives larger $J/\psi J/\psi\gamma$ cross sections for all four phase choices, as expected from its larger electromagnetic transition coupling. The change of model affects the normalization more strongly than the qualitative phase ordering: $\phi_X=0$ gives the largest near-resonance rate in both cases, while $\phi_X=\pi$ gives the smallest. The phase dependence is therefore not tied to a specific normalization of $g_{T\gamma}$.

For $\eta_c\eta_c\gamma$, the full-region cross section remains close to $1.4\times10^{-2}~{\rm fb}$ in both models. Across the four phase choices, it varies by about $4\%$, reflecting the dominance of the continuum in the total rate. The near-resonance cross section is more sensitive to the phase, ranging from approximately
$1.90\times10^{-3}$ to $2.08\times10^{-3}~{\rm fb}$ in Model I and reaching $2.09\times10^{-3}~{\rm fb}$ in Model II. These variations are at the level of approximately $9$--$10\%$, consistent with a localized interference contribution superimposed on a
continuum-dominated distribution.

The $J/\psi J/\psi\gamma$ channel thus retains a strong dependence on both the electromagnetic transition coupling and the relative phase. By contrast, the $\eta_c\eta_c\gamma$ rate is dominated by the
continuum, and the tensor contribution is reflected mainly in the local distribution near $M_X$.

\section{Summary}

We have investigated the contribution of a tensor $X(6900)$ to the radiative double-charmonium processes $e^+e^-\to J/\psi J/\psi\gamma$ and $e^+e^-\to\eta_c\eta_c\gamma$ at $\sqrt{s}=10.58~{\rm GeV}$. The nonresonant amplitudes were calculated within color-singlet NRQCD framework, while the resonance contribution was described using effective electromagnetic and hadronic couplings whose magnitudes were inferred from the corresponding partial decay widths. Two representative predictions for $\Gamma(X\to\gamma\gamma)$ were adopted to assess the dependence on the electromagnetic input.

The two final states exhibit distinct manifestations of the tensor resonance. In the $J/\psi J/\psi\gamma$ channel, the $X(6900)$ produces a pronounced structure near its nominal mass, and interference with the continuum substantially modifies both its magnitude and local line shape. In the $\eta_c\eta_c\gamma$ channel, the distribution is dominated by the NRQCD continuum, while the resonance appears primarily through a localized interference effect near
$M_{\eta_c\eta_c}\simeq M_X$. The difference between the two channels arises from their distinct effective vertices and their contractions with the corresponding continuum amplitudes.

The contribution of the $X(6900)$ therefore cannot, in general, be represented by an incoherent Breit--Wigner term added to the continuum spectrum. A consistent description requires the continuum and resonance amplitudes to be combined coherently. Radiative double-charmonium production thus provides a complementary avenue for examining the channel-dependent line-shape features of the $X(6900)$ and for constraining its effective electromagnetic and hadronic couplings.
Although direct observation of the interference-modified line shape of $(X(6900))$ is statistically challenging for Belle II, these predictions provide reliable theoretical benchmarks for future high-luminosity $e^+e^-$ collider experiments.

\section*{Acknowledgements} 
This work was supported by the National Natural Science Foundation of China (No. 11705078, 12575087).

\section*{Appendix: Continuum invariant-mass differential cross sections}

We consider the continuum process
\begin{equation}
	e^-(l_1)+e^+(l_2)\to \gamma^\ast(k)
	\to H(p_3)+H(p_4)+\gamma(p_5),
	\qquad H=J/\psi,\eta_c .
\end{equation}
we introduce the dimensionless variables
$z_3=2E_3/\sqrt{s}, r=4m_c^2 / s,$
with
\begin{equation}
	dE_3=\frac{\sqrt{s}}{2}\,dz_3,
	\qquad
	\frac{M_{HH}}{\sqrt{s}}\,dE_3
	=\frac{M_{HH}}{2}\,dz_3 .
\end{equation}
The dimensionless integration limits are
$z_3^{\min,\max}=2E_3^{\min,\max}/\sqrt{s}$.
the continuum invariant-mass differential cross sections are given by

\subsection{The $J/\psi J/\psi\gamma$ channel}
For $H=J/\psi$, the continuum differential cross section is
\begingroup
\allowdisplaybreaks[4]
\setlength{\jot}{1.5pt}
\begin{flalign*}
	\frac{d\sigma_{\rm cont}^{J/\psi}}{dM_{J/\psi J/\psi}dz_{3}}
	&={}\frac{1}{2!}\,\frac{1}{4}\,\frac{1}{s^{2}}\left(\frac{4s}{3}\,4\pi\alpha\right)\frac{1}{2s} \, \frac{1}{4(2\pi)^{3}}\,\frac{M_{J/\psi J/\psi}}{2}\, && \\
	&{}\times\,268435456 \pi^{5} \alpha^{3} \alpha_{s}^{2} \bigl(64 M_{J/\psi J/\psi}^{20} r^{2} - 64 M_{J/\psi J/\psi}^{18} r s \bigl(r^{2} + r \bigl(6 z_{3} + 5\bigr) - 20 z_{3} && \\
	&{}+\,21\bigr) + 4 M_{J/\psi J/\psi}^{16} s^{2} \bigl(3 r^{4} + r^{3} \bigl(336 - 88 z_{3}\bigr) + 4 r^{2} \bigl(20 z_{3}^{2} + 368 z_{3} - 241\bigr) && \\
	&{}-\,16 r \bigl(92 z_{3}^{2} + 100 z_{3} - 201\bigr) + 800 z_{3}^{2} - 1664 z_{3} + 864\bigr) + 8 M_{J/\psi J/\psi}^{14} s^{3} \bigl(3 r^{4} \bigl(2 z_{3} && \\
	&{}-\,9\bigr) + 2 r^{3} \bigl(3 z_{3}^{2} + 255 z_{3} - 670\bigr) + 2 r^{2} \bigl(56 z_{3}^{3} - 592 z_{3}^{2} - 1826 z_{3} + 2495\bigr) && \\
	&{}+\,6 r \bigl(144 z_{3}^{3} + 1088 z_{3}^{2} - 172 z_{3} - 1103\bigr) - 1568 z_{3}^{3} - 192 z_{3}^{2} + 5472 z_{3} - 3712\bigr) && \\
	&{}-\,4 M_{J/\psi J/\psi}^{12} s^{4} \bigl(6 r^{4} \bigl(21 z_{3} - 62\bigr) - 4 r^{3} \bigl(38 z_{3}^{3} - 187 z_{3}^{2} - 1063 z_{3} + 2761\bigr) && \\
	&{}+\,r^{2} \bigl(496 z_{3}^{4} - 528 z_{3}^{3} - 13476 z_{3}^{2} - 17236 z_{3} + 40383\bigr) - 8 r \bigl(104 z_{3}^{4} - 2868 z_{3}^{3} && \\
	&{}-\,4472 z_{3}^{2} + 3299 z_{3} + 4044\bigr) - 5664 z_{3}^{4} - 12544 z_{3}^{3} + 19328 z_{3}^{2} + 25616 z_{3} - 26736\bigr) && \\
	&{}-\,4 M_{J/\psi J/\psi}^{10} s^{5} \bigl(3 r^{4} \bigl(8 z_{3}^{3} - 34 z_{3}^{2} - 136 z_{3} + 425\bigr) + r^{3} \bigl(76 z_{3}^{4} + 920 z_{3}^{3} && \\
	&{}-\,5191 z_{3}^{2} - 7634 z_{3} + 26404\bigr) - 4 r^{2} \bigl(48 z_{3}^{5} + 756 z_{3}^{4} - 2814 z_{3}^{3} - 7423 z_{3}^{2} - 6090 z_{3} && \\
	&{}+\,23593\bigr) + 4 r \bigl(688 z_{3}^{5} - 1720 z_{3}^{4} - 23952 z_{3}^{3} - 2841 z_{3}^{2} + 13678 z_{3} + 14227\bigr) + 5632 z_{3}^{5} && \\
	&{}+\,28800 z_{3}^{4} - 3840 z_{3}^{3} - 58416 z_{3}^{2} - 26672 z_{3} + 54496\bigr) + M_{J/\psi J/\psi}^{8} s^{6} \bigl(3 r^{4} \bigl(16 z_{3}^{4} && \\
	&{}+\,128 z_{3}^{3} - 724 z_{3}^{2} - 652 z_{3} + 3261\bigr) + 8 r^{3} \bigl(328 z_{3}^{4} + 605 z_{3}^{3} - 6893 z_{3}^{2} - 2210 z_{3} && \\
	&{}+\,19319\bigr) + 8 r^{2} \bigl(192 z_{3}^{6} - 2136 z_{3}^{5} + 1770 z_{3}^{4} + 10838 z_{3}^{3} + 20526 z_{3}^{2} + 9734 z_{3} && \\
	&{}-\,69847\bigr) + 32 r \bigl(264 z_{3}^{6} + 1140 z_{3}^{5} - 6830 z_{3}^{4} - 22011 z_{3}^{3} + 14891 z_{3}^{2} + 2127 z_{3} && \\
	&{}+\,10263\bigr) + 12288 z_{3}^{6} + 123136 z_{3}^{5} + 140544 z_{3}^{4} - 280000 z_{3}^{3} - 209792 z_{3}^{2} - 76288 z_{3} && \\
	&{}+\,290112\bigr) - 4 M_{J/\psi J/\psi}^{6} s^{7} \bigl(3 r^{4} \bigl(20 z_{3}^{4} + 34 z_{3}^{3} - 369 z_{3}^{2} + 4 z_{3} + 916\bigr) + r^{3} \bigl(144 z_{3}^{5} && \\
	&{}+\,1261 z_{3}^{4} - 554 z_{3}^{3} - 17908 z_{3}^{2} + 4078 z_{3} + 35205\bigr) + 4 r^{2} \bigl(96 z_{3}^{7} - 260 z_{3}^{6} - 1992 z_{3}^{5} && \\
	&{}+\,6137 z_{3}^{4} - 529 z_{3}^{3} + 15240 z_{3}^{2} - 1699 z_{3} - 32663\bigr) + 4 r \bigl(256 z_{3}^{7} + 1512 z_{3}^{6} + 2952 z_{3}^{5} && \\
	&{}-\,37411 z_{3}^{4} - 26226 z_{3}^{3} + 52932 z_{3}^{2} - 17950 z_{3} + 23332\bigr) + 1024 z_{3}^{7} + 14528 z_{3}^{6} && \\
	&{}+\,61632 z_{3}^{5} - 52144 z_{3}^{4} - 47312 z_{3}^{3} - 34304 z_{3}^{2} - 7424 z_{3} + 64000\bigr) && \\
	&{}+\,2 M_{J/\psi J/\psi}^{4} s^{8} \bigl(3 r^{4} \bigl(74 z_{3}^{4} - 28 z_{3}^{3} - 710 z_{3}^{2} + 344 z_{3} + 1195\bigr) + 4 r^{3} \bigl(24 z_{3}^{6} && \\
	&{}+\,144 z_{3}^{5} + 267 z_{3}^{4} - 651 z_{3}^{3} - 6434 z_{3}^{2} + 3849 z_{3} + 9771\bigr) + 2 r^{2} \bigl(96 z_{3}^{8} + 768 z_{3}^{7} && \\
	&{}-\,5376 z_{3}^{6} + 2892 z_{3}^{5} + 20180 z_{3}^{4} - 22716 z_{3}^{3} + 68064 z_{3}^{2} - 32316 z_{3} - 72791\bigr) && \\
	&{}+\,16 r \bigl(32 z_{3}^{8} + 352 z_{3}^{7} + 520 z_{3}^{6} + 1720 z_{3}^{5} - 27862 z_{3}^{4} + 17049 z_{3}^{3} + 7678 z_{3}^{2} && \\
	&{}-\,7986 z_{3} + 8246\bigr) + 512 z_{3}^{8} + 6144 z_{3}^{7} + 58496 z_{3}^{6} + 72928 z_{3}^{5} - 268800 z_{3}^{4} + 211104 z_{3}^{3} && \\
	&{}-\,216192 z_{3}^{2} + 77696 z_{3} + 58112\bigr) - 4 M_{J/\psi J/\psi}^{2} s^{9} \bigl(3 r^{4} \bigl(30 z_{3}^{4} - 42 z_{3}^{3} - 157 z_{3}^{2} && \\
	&{}+\,136 z_{3} + 209\bigr) + r^{3} \bigl(144 z_{3}^{6} - 108 z_{3}^{5} + 33 z_{3}^{4} - 246 z_{3}^{3} - 5461 z_{3}^{2} + 4612 z_{3} && \\
	&{}+\,6025\bigr) + 4 r^{2} \bigl(72 z_{3}^{8} - 72 z_{3}^{7} - 963 z_{3}^{6} + 2277 z_{3}^{5} - 421 z_{3}^{4} - 1825 z_{3}^{3} + 8638 z_{3}^{2} && \\
	&{}-\,6142 z_{3} - 5357\bigr) + 4 r \bigl(240 z_{3}^{8} - 96 z_{3}^{7} + 2224 z_{3}^{6} - 2088 z_{3}^{5} - 35999 z_{3}^{4} + 53130 z_{3}^{3} && \\
	&{}-\,26767 z_{3}^{2} + 4388 z_{3} + 4758\bigr) + 1024 z_{3}^{8} + 128 z_{3}^{7} + 39760 z_{3}^{6} - 54544 z_{3}^{5} - 13680 z_{3}^{4} && \\
	&{}+\,72144 z_{3}^{3} - 97376 z_{3}^{2} + 51392 z_{3} + 1152\bigr) + s^{10} \bigl(3 r^{4} \bigl(- 6 z_{3}^{2} + 6 z_{3} + 11\bigr)^{2} && \\
	&{}+\,8 r^{3} \bigl(54 z_{3}^{6} - 162 z_{3}^{5} + 180 z_{3}^{4} - 90 z_{3}^{3} - 495 z_{3}^{2} + 513 z_{3} + 391\bigr) + 32 r^{2} \bigl(27 z_{3}^{8} && \\
	&{}-\,108 z_{3}^{7} + 108 z_{3}^{6} + 54 z_{3}^{5} - 39 z_{3}^{4} - 138 z_{3}^{3} + 878 z_{3}^{2} - 782 z_{3} - 301\bigr) + 96 r \bigl(36 z_{3}^{8} && \\
	&{}-\,144 z_{3}^{7} + 572 z_{3}^{6} - 1212 z_{3}^{5} - 169 z_{3}^{4} + 2190 z_{3}^{3} - 2101 z_{3}^{2} + 828 z_{3} - 6\bigr) && \\
	&{}+\,128 z_{3} \bigl(33 z_{3}^{7} - 132 z_{3}^{6} + 987 z_{3}^{5} - 2499 z_{3}^{4} + 3134 z_{3}^{3} - 2257 z_{3}^{2} + 698 z_{3} + 36\bigr)\bigr)\bigr) && \\
	&{}/\,\Bigl\{ 2187 r s^{6} \bigl(- 2 M_{J/\psi J/\psi}^{2} + s \bigl(z_{3} + 2\bigr)\bigr)^{2} \bigl(M_{J/\psi J/\psi}^{2} - s z_{3}\bigr)^{2} \bigl(M_{J/\psi J/\psi}^{2} + s \bigl(z_{3} && \\
	&{}-\,3\bigr)\bigr)^{2} \bigl(2 M_{J/\psi J/\psi}^{2} - 3 s\bigr)^{2} \bigl(z_{3} - 1\bigr)^{2}\Bigr\}. && \\
\end{flalign*}

\subsection{The $\eta_c\eta_c\gamma$ channel}
For $H=\eta_c$, the continuum differential cross section is
\begingroup
\allowdisplaybreaks[4]
\setlength{\jot}{1.5pt}
\begin{flalign*}
	\frac{d\sigma_{\rm cont}^{\eta_c}}{dM_{\eta_c\eta_c} dz_{3}} 
	&={}\frac{1}{2!}\,\frac{1}{4}\,\frac{1}{s^{2}}\left(\frac{4s}{3}\,4\pi\alpha\right)\frac{1}{2s} \, \frac{1}{4(2\pi)^{3}}\,\frac{M_{\eta_c\eta_c}}{2}
	\, && \\
	&{}\times\,268435456 \pi^{5} \alpha^{3} \alpha_{s}^{2} \bigl(M_{\eta_c\eta_c}^{16} \bigl(r^{4} + 4 r^{3} \bigl(2 z_{3} - 5\bigr) + 4 r^{2} \bigl(4 z_{3}^{2} - 16 z_{3} + 21\bigr) && \\
	&{}+\,16 r \bigl(z_{3}^{2} - 6 z_{3} + 4\bigr) + 128 z_{3}^{4} - 832 z_{3}^{3} + 1920 z_{3}^{2} - 1792 z_{3} + 576\bigr) && \\
	&{}-\,4 M_{\eta_c\eta_c}^{14} s \bigl(r^{4} \bigl(z_{3} + 1\bigr) + 4 r^{3} \bigl(z_{3}^{2} - 7\bigr) + 12 r^{2} \bigl(4 z_{3}^{2} - 13 z_{3} + 15\bigr) + 8 r \bigl(4 z_{3}^{4} && \\
	&{}-\,15 z_{3}^{3} + 7 z_{3}^{2} + 4 z_{3} - 4\bigr) + 128 z_{3}^{5} - 752 z_{3}^{4} + 1328 z_{3}^{3} - 288 z_{3}^{2} - 928 z_{3} + 512\bigr) && \\
	&{}+\,2 M_{\eta_c\eta_c}^{12} s^{2} \bigl(r^{4} \bigl(4 z_{3}^{2} + 8 z_{3} - 1\bigr) + 2 r^{3} \bigl(21 z_{3}^{2} - 50 z_{3} - 45\bigr) - 4 r^{2} \bigl(6 z_{3}^{4} && \\
	&{}-\,66 z_{3}^{3} + 54 z_{3}^{2} + 178 z_{3} - 291\bigr) + 8 r \bigl(32 z_{3}^{5} - 111 z_{3}^{4} + 26 z_{3}^{3} + 96 z_{3}^{2} - 10 z_{3} && \\
	&{}-\,59\bigr) + 384 z_{3}^{6} - 2016 z_{3}^{5} + 2304 z_{3}^{4} + 2464 z_{3}^{3} - 3392 z_{3}^{2} - 1440 z_{3} + 1696\bigr) && \\
	&{}-\,4 M_{\eta_c\eta_c}^{10} s^{3} \bigl(r^{4} \bigl(2 z_{3}^{3} + 9 z_{3}^{2} - 6\bigr) - 2 r^{3} \bigl(5 z_{3}^{4} - 19 z_{3}^{3} + 4 z_{3}^{2} + 79 z_{3} - 15\bigr) && \\
	&{}+\,4 r^{2} \bigl(2 z_{3}^{5} + 9 z_{3}^{4} + 61 z_{3}^{3} - 166 z_{3}^{2} - 21 z_{3} + 220\bigr) + 8 r \bigl(24 z_{3}^{6} - 64 z_{3}^{5} && \\
	&{}-\,69 z_{3}^{4} + 183 z_{3}^{3} - 18 z_{3}^{2} - 21 z_{3} - 57\bigr) + 128 z_{3}^{7} - 592 z_{3}^{6} + 240 z_{3}^{5} + 1328 z_{3}^{4} && \\
	&{}+\,656 z_{3}^{3} - 3024 z_{3}^{2} + 624 z_{3} + 640\bigr) + M_{\eta_c\eta_c}^{8} s^{4} \bigl(r^{4} \bigl(4 z_{3}^{4} + 40 z_{3}^{3} + 28 z_{3}^{2} - 64 z_{3} && \\
	&{}-\,19\bigr) - 4 r^{3} \bigl(12 z_{3}^{5} + z_{3}^{4} - 142 z_{3}^{3} + 240 z_{3}^{2} + 128 z_{3} - 172\bigr) + 4 r^{2} \bigl(40 z_{3}^{6} - 28 z_{3}^{5} && \\
	&{}+\,60 z_{3}^{4} + 12 z_{3}^{3} - 648 z_{3}^{2} + 304 z_{3} + 693\bigr) + 16 r \bigl(32 z_{3}^{7} - 32 z_{3}^{6} - 336 z_{3}^{5} + 465 z_{3}^{4} && \\
	&{}+\,134 z_{3}^{3} - 177 z_{3}^{2} - 12 z_{3} - 115\bigr) + 128 z_{3}^{8} - 512 z_{3}^{7} - 256 z_{3}^{6} + 256 z_{3}^{5} + 7424 z_{3}^{4} && \\
	&{}-\,6144 z_{3}^{3} - 5632 z_{3}^{2} + 3840 z_{3} + 896\bigr) - 4 M_{\eta_c\eta_c}^{6} s^{5} \bigl(r^{4} \bigl(6 z_{3}^{4} + 14 z_{3}^{3} - 19 z_{3}^{2} && \\
	&{}-\,15 z_{3} + 7\bigr) - 2 r^{3} \bigl(2 z_{3}^{6} + 18 z_{3}^{5} - 60 z_{3}^{4} - 29 z_{3}^{3} + 216 z_{3}^{2} - 55 z_{3} - 94\bigr) && \\
	&{}+\,4 r^{2} \bigl(8 z_{3}^{7} + 17 z_{3}^{6} - 79 z_{3}^{5} + 49 z_{3}^{4} + 19 z_{3}^{3} - 87 z_{3}^{2} + 71 z_{3} + 71\bigr) + 8 r \bigl(4 z_{3}^{8} && \\
	&{}+\,16 z_{3}^{7} - 128 z_{3}^{6} + 92 z_{3}^{5} + 156 z_{3}^{4} - 68 z_{3}^{3} - 79 z_{3}^{2} + 37 z_{3} - 40\bigr) - 64 z_{3} \bigl(3 z_{3}^{5} && \\
	&{}-\,9 z_{3}^{4} - 4 z_{3}^{3} + 14 z_{3}^{2} + 4 z_{3} - 8\bigr)\bigr) + 4 M_{\eta_c\eta_c}^{4} s^{6} \bigl(r^{4} \bigl(13 z_{3}^{4} - 2 z_{3}^{3} - 33 z_{3}^{2} && \\
	&{}+\,8 z_{3} + 12\bigr) + r^{3} \bigl(- 16 z_{3}^{6} + 159 z_{3}^{4} - 174 z_{3}^{3} - 189 z_{3}^{2} + 156 z_{3} + 88\bigr) + 8 r^{2} \bigl(z_{3}^{8} && \\
	&{}+\,12 z_{3}^{7} - 28 z_{3}^{6} - 28 z_{3}^{5} + 83 z_{3}^{4} - 40 z_{3}^{3} - 3 z_{3}^{2} + 12 z_{3} + 6\bigr) + 16 r \bigl(4 z_{3}^{8} && \\
	&{}-\,16 z_{3}^{7} - 2 z_{3}^{6} + 26 z_{3}^{5} + 44 z_{3}^{4} - 78 z_{3}^{3} + 25 z_{3}^{2} + 3 z_{3} - 7\bigr) + 288 z_{3}^{2} \bigl(z_{3} - 1\bigr)^{2}\bigr) && \\
	&{}-\,8 M_{\eta_c\eta_c}^{2} r s^{7} \bigl(r^{3} \bigl(6 z_{3}^{4} - 8 z_{3}^{3} - 7 z_{3}^{2} + 7 z_{3} + 3\bigr) - 2 r^{2} \bigl(4 z_{3}^{6} - 12 z_{3}^{5} && \\
	&{}+\,2 z_{3}^{4} + 14 z_{3}^{3} + 9 z_{3}^{2} - 15 z_{3} - 4\bigr) + 4 r \bigl(4 z_{3}^{8} - 42 z_{3}^{6} + 62 z_{3}^{5} - 8 z_{3}^{4} - 26 z_{3}^{3} && \\
	&{}+\,19 z_{3}^{2} - 5 z_{3} + 1\bigr) - 48 z_{3} \bigl(2 z_{3}^{5} - 6 z_{3}^{4} + 4 z_{3}^{3} + 2 z_{3}^{2} - 3 z_{3} + 1\bigr)\bigr) && \\
	&{}+\,4 r^{2} s^{8} \bigl(r^{2} \bigl(- 2 z_{3}^{2} + 2 z_{3} + 1\bigr)^{2} + 4 r z_{3} \bigl(z_{3}^{3} - 2 z_{3}^{2} - z_{3} + 2\bigr) + 8 \bigl(- 2 z_{3}^{4} && \\
	&{}+\,4 z_{3}^{3} - 2 z_{3} + 1\bigr)^{2}\bigr)\bigr) && \\
	&{}/\,\Bigl\{ 2187 M_{\eta_c\eta_c}^{8} r s^{2} \bigl(M_{\eta_c\eta_c}^{2} - 2 s\bigr)^{2} \bigl(M_{\eta_c\eta_c}^{2} - s z_{3}\bigr)^{2} \bigl(M_{\eta_c\eta_c}^{2}-\,s \bigl(z_{3} + 1\bigr)\bigr)^{2} \bigl(z_{3} - 2\bigr)^{2} \bigl(z_{3} - 1\bigr)^{2}\Bigr\}. && \\
\end{flalign*}

\end{document}